# Resilience of Task-Oriented V2X Networks to Incomplete Information Sharing

Luca Lusvarghi, *Member, IEEE* and Javier Gozalvez, *Fellow, IEEE*

***Abstract*—Task-oriented Vehicle-to-Everything (V2X) networks have recently been proposed to scalably support the large-scale deployment of connected vehicles. In task-oriented V2X networks, vehicles select the content of the transmitted messages based on its relevance to the intended receivers. However, estimating relevance can be challenging, especially in highly dynamic and complex driving scenarios. Relevance estimation errors may cause a transmitting vehicle to share incomplete information, omitting relevant data that is critical for the intended receivers' situational awareness. This work numerically demonstrates that task-oriented V2X networks exhibit an inherent resilience to incomplete information sharing. We show that such resilience guarantees a consistent delivery of relevant information even under high relevance estimation error probability conditions. Furthermore, we show that the fundamental conditions underpinning such inherent resilience can also be encountered outside of the V2X domain – in particular, in other task-oriented networks where multiple transmitters select the content of their messages based on the task-related requirements of a common set of intended receivers.**

## I. INTRODUCTION

Vehicle-to-Everything (V2X) communications are a key enabler for the realization of safer and more efficient forms of connected and automated mobility. Through V2X, connected vehicles can cooperatively enhance their situational awareness and make more informed driving decisions by exchanging a broad range of information, such as their position, speed, and detected objects.

Realizing the full potential of V2X connectivity requires the large-scale deployment of connected vehicles. However, the scalable support of a large number of connected vehicles on V2X networks is currently challenged by the increasing connectivity demand of cooperative driving functionalities and the limitations of existing V2X communication systems. Existing systems primarily focus on the timely and reliable delivery of the transmitted messages, without fully considering the usefulness of the transmitted information to the intended receivers. As the number of connected vehicles grows, this approach risks saturating the available network resources [1].

Recent contributions have demonstrated that task-oriented communications can greatly improve the scalability of future networks [2]. In task-oriented networks, connected devices transmit only the information required by the tasks executed at the intended receivers, thereby improving communication efficiency and reducing the consumption of network resources. The potential of task-oriented communications has also been explored in the V2X domain to address the scalability challenges of future V2X networks [3]-[6]. In particular, task-oriented V2X networks where vehicles curate the transmitted information based on its relevance to the intended receivers have been recently proposed in [7]. In task-oriented V2X networks, the relevance of a piece of information captures its impact on the intended receiver's driving: information is deemed relevant if it influences the driving of the receiver and irrelevant if it does not. Preliminary results reported in [7] show that, by focusing on relevance, vehicles can substantially improve communication efficiency and network scalability compared to existing V2X communication approaches [8].

The improved scalability of task-oriented V2X networks relies on the capability of the transmitting vehicles to identify the information that is relevant to the intended receivers – and intelligently curate the content of the transmitted messages to avoid the transmission of unnecessary (or irrelevant) data. Yet, relevance estimation can be particularly challenging in the V2X domain, mainly due to the highly dynamic and complex nature of the driving environment. For example, the relevance of an object (e.g., a vehicle or vulnerable road user) can rapidly vary over time and depends on its interactions with other surrounding objects. Relevance estimation is even more challenging due to the broadcast nature of V2X communications, as each transmitting device needs to serve multiple intended receivers with distinct driving requirements. As a result, the same piece of information might have a different relevance to each intended receiver. An inaccurate relevance estimation may cause a transmitting vehicle to omit relevant information required by one or multiple receivers from the transmitted message, causing a Content-Selection Error (CSE). Content-selection errors lead to the transmission of incomplete information and, therefore, can impair the receiving vehicles' situational awareness, potentially affecting their driving safety and efficiency. For example, a CSE could prevent the transmission of information about a relevant object crossing the receiver's trajectory and potentially representing a safety-critical risk.

In this work, we numerically demonstrate that task-oriented V2X networks feature an inherent resilience to content-selection errors: when a transmitting vehicle wrongly omits a

This work was supported by the European Union under the 2023 MSCA Postdoctoral Fellowship program (project no. 101153845). This work was also partially funded by MICIU/AEI/10.13039/501100011033, "ERFD/EU" (PID2023-150308OB-I00), and Generalitat Valenciana (CIAICO/2024/167). *(Corresponding author: Luca Lusvarghi)*.

Luca Lusvarghi and Javier Gozalvez are with the Networked Systems Lab, Universidad Miguel Hernandez de Elche, Elche, 03202, Spain. E-mail: {llusvarghi, j.gozalvez}@umh.es.

piece of relevant information, neighbouring vehicles naturally intervene to transmit the omitted information and compensate for the content-selection error. Simulation results show that this resilience guarantees a consistent delivery of relevant information at the receiving vehicles also in high content-selection error probability scenarios, thereby relaxing the need for very accurate relevance estimations. Results also show that this resilience is stronger when the number of connected vehicles is larger, as more vehicles can potentially compensate for each other's content-selection errors. Furthermore, our analysis identifies the fundamental conditions that underpin the inherent resilience of task-oriented V2X networks, revealing that these conditions can also be encountered outside of the V2X domain. In particular, the analysis shows that these conditions could potentially be encountered in other task-oriented networks where multiple transmitting devices select the content of their messages based on the task-related requirements of a common set of intended receivers.

The rest of this paper is organized as follows. Section II reviews the state-of-the-art and positions our contribution. Section III sheds light on the key communication principles, as well as on the main content-selection error causes, that characterize task-oriented V2X networks. Section IV introduces the cooperative perception case study employed for our analysis and Section V describes the evaluation scenario. Section VI presents the obtained numerical results and Section VII draws the conclusions.

## II. Related Work

Task-oriented networks have recently emerged as a promising solution to cope with the exponentially increasing connectivity demand that will characterize the 6G and beyond era. As of today, the enhanced scalability of task-oriented networks has already been demonstrated in a wide range of Internet-of-Things (IoT) use cases. For instance, a task-oriented network was employed in [9] in the context of a multi-user Metaverse construction task, where distributed cameras share semantically segmented images required to virtualize each user's environment. The authors put forth a scheduling algorithm that selectively forwards to each user only the images required to virtualize its surrounding environment. This approach substantially reduces the network load and the amount of unnecessary information processed by each user with respect to existing task-agnostic approaches, while simultaneously enhancing the overall Quality of Experience (QoE). The work in [10] explored the potential of task-oriented networks in the context of a multi-view inference task executed at the edge. By jointly optimizing the selection of the image features transmitted by distributed cameras, the proposed approach minimizes the amount of exchanged data without affecting the inference accuracy. A similar approach is put forth in [11] considering a radar-based human motion recognition task. With respect to existing approaches, a task-oriented network maximizes inference accuracy and meets real-time latency constraints while minimizing the network load.

Task-oriented networks have also been applied to the cooperative perception use case. In cooperative perception, intelligent agents (e.g., robots, drones, or autonomous vehicles) share raw or processed (detected objects) sensing data to overcome their onboard sensors limitations and collaboratively achieve a longer range and more accurate perception of the surrounding environment. A seminal work applying pull-based task-oriented networks to cooperative perception is [12]. In [12], the authors put forth a learnable handshaking mechanism that enables Unmanned Aerial Vehicles (UAVs) to pull relevant sensing data from a selected subset of neighbouring UAVs to improve the accuracy of a local semantic segmentation task. Results show that the proposed handshake mechanism achieves a superior semantic segmentation performance compared to baseline solutions while consuming less network resources. The work in [12] is extended in [13], where the handshaking mechanism is enhanced with a scheduling policy that pulls additional data only when the accuracy of local data is insufficient to meet a predefined accuracy level. A further extension to the works in [12] and [13], tailored to an object detection task, is represented by [14]. In [14], an intelligent agent pulls additional raw sensing data only from neighbouring agents that have detected relevant objects. The relevance of an object is defined as a function of its redundancy and proximity relative to the requesting agent. The proposed solution is evaluated in the V2X domain, where the intelligent agents are connected vehicles that exchange data via V2X communications. Within the V2X domain, task-oriented networks are commonly referred to as task-oriented V2X networks. The results in [14] show that, through a pull-based task-oriented V2X network, vehicles can reliably detect the most relevant objects at the minimum communication cost.

State-of-the-art works employing push-based task-oriented V2X networks to support cooperative perception are [15] and [16]. In [15], vehicles broadcast a selected subset of region-level features extracted from raw sensing data (e.g., images and pointclouds) regions that contain objects. By avoiding the transmission of regions with background information, the proposed approach reduces the consumption of network resources without compromising the performance of camera- and LiDAR-based 3D object detection tasks. The approach proposed in [15] is refined by the authors of [16]. In [16], vehicles transmit object-level sparse features instead of region-level features. By focusing on the extraction of object-level information, the proposed approach further reduces the amount of transmitted data and simultaneously enhances the accuracy of LiDAR-based 3D object detection tasks.

The transmission of object-level information (e.g., object position and speed) is also envisioned by V2X cooperative perception standards [8]. Current standards establish that vehicles should exchange the objects detected in raw sensing data locally collected through their onboard sensors. In standardized cooperative perception, a detected object can represent any type of road user or obstacle in the driving environment (e.g., vehicles, pedestrians, bicycles, traffic cones). Task-oriented V2X networks tailored to standard-compliant cooperative perception have been put forth in [3]-[6]. Their objective is to limit the number of detected objects transmitted by each vehicle while maximizing the performance of the cooperative perception task. In [3] and [4], the subset of transmitted objects is respectively selected based on the objects' entropy at the receiver and the rate of their position, speed, and heading variations. In [5] and [6], transmitted objects are selected based on their detection uncertainty and accuracy, respectively. Other proposals based on the objects redundancy and classification confidence have also been included in the cooperative perception standard [8]. More recently, task-oriented V2X networks where vehicles curate the subset of transmitted objects based on its relevance to the intended receivers were proposed in [7]. In [7], cooperative perception is not regarded as the goal of the communication, but as a means to plan more informed driving decisions. Hence, the relevance of each detected object is contextualized with respect to the driving requirements of each intended receiver. A detected object is deemed relevant if it can influence the driving of the receiving vehicle and considered irrelevant otherwise. The results reported in [7] show that task-oriented V2X networks based on the relevance of the exchanged objects can achieve a twofold improvement in terms of communication efficiency with respect to other existing task-oriented proposals.

The effectiveness of task-oriented networks depends on the ability of the transmitting devices to accurately select the content of the generated messages and transmit only the information required by each intended receiver's tasks – a critical yet largely unexplored challenge. To date, content-selection challenges have only been preliminarily investigated in the context of task-oriented V2X networks by [17]. The results in [17] show that vehicles can make a content-selection error and, therefore, share incomplete information once every five transmitted messages, on average. In the cooperative perception context, a content-selection error occurs when the transmitter wrongly omits at least one relevant object from the transmitted message due to an inaccurate estimation of the objects' relevance to the intended receivers. Content-selection errors can be particularly critical as they reduce the effectiveness of task-oriented V2X networks and impair the receiving vehicles' situational awareness.

## III. Task-Oriented V2X Networks

In task-oriented V2X networks, transmitting vehicles select the content of the transmitted messages based on its relevance to the intended receivers [7]. Relevance captures the impact that a piece of information has on the receiving vehicle's driving and it can greatly vary depending on context, especially in the V2X domain. For example, in cooperative perception, an object crossing an intersection is relevant for a vehicle heading towards the same intersection, as it can potentially represent a safety-critical situation. In contrast, the same object is irrelevant for a vehicle leaving the intersection. Context refers to the driving and communication environment in which a piece of information is exchanged or processed. It is defined by the receiver's current and future state (position, heading, and speed), previously exchanged V2X messages, and the mobility of surrounding objects, for example.

In a task-oriented V2X network, each transmitting vehicle needs to estimate the context-dependent relevance of locally collected information[1] (e.g., detected objects) for its intended receivers to optimally select the content of the transmitted messages. To do so, a transmitting vehicle first performs redundancy estimation to determine what information is already available at the intended receivers. Redundant information is assumed to not improve the intended receivers' understanding of the driving environment and, therefore, to not have any impact on their driving. For this reason, information estimated to be redundant is considered irrelevant and is not transmitted[2]. Then, it estimates the relevance of the remaining non-redundant locally collected information. Relevance estimation consists of identifying what information could have an impact on the driving of each intended receiver, based on its context. Due to the broadcast nature of V2X communications, a transmitting vehicle typically needs to estimate the relevance of each piece of information considering multiple intended receivers. The resulting multi-receiver relevance estimates are leveraged to select the content of the generated messages. The content-selection goal is to exclude from the generated message all the information that is irrelevant for all intended receivers. Note that a transmitting vehicle may drop a generated message prior to its transmission based on the indications from the congestion control mechanism. In V2X, congestion control is designed to keep the transmitting vehicle's consumption of network resources below a pre-defined threshold that depends on the network load. The operation of a transmitting vehicle is illustrated in Fig. 1

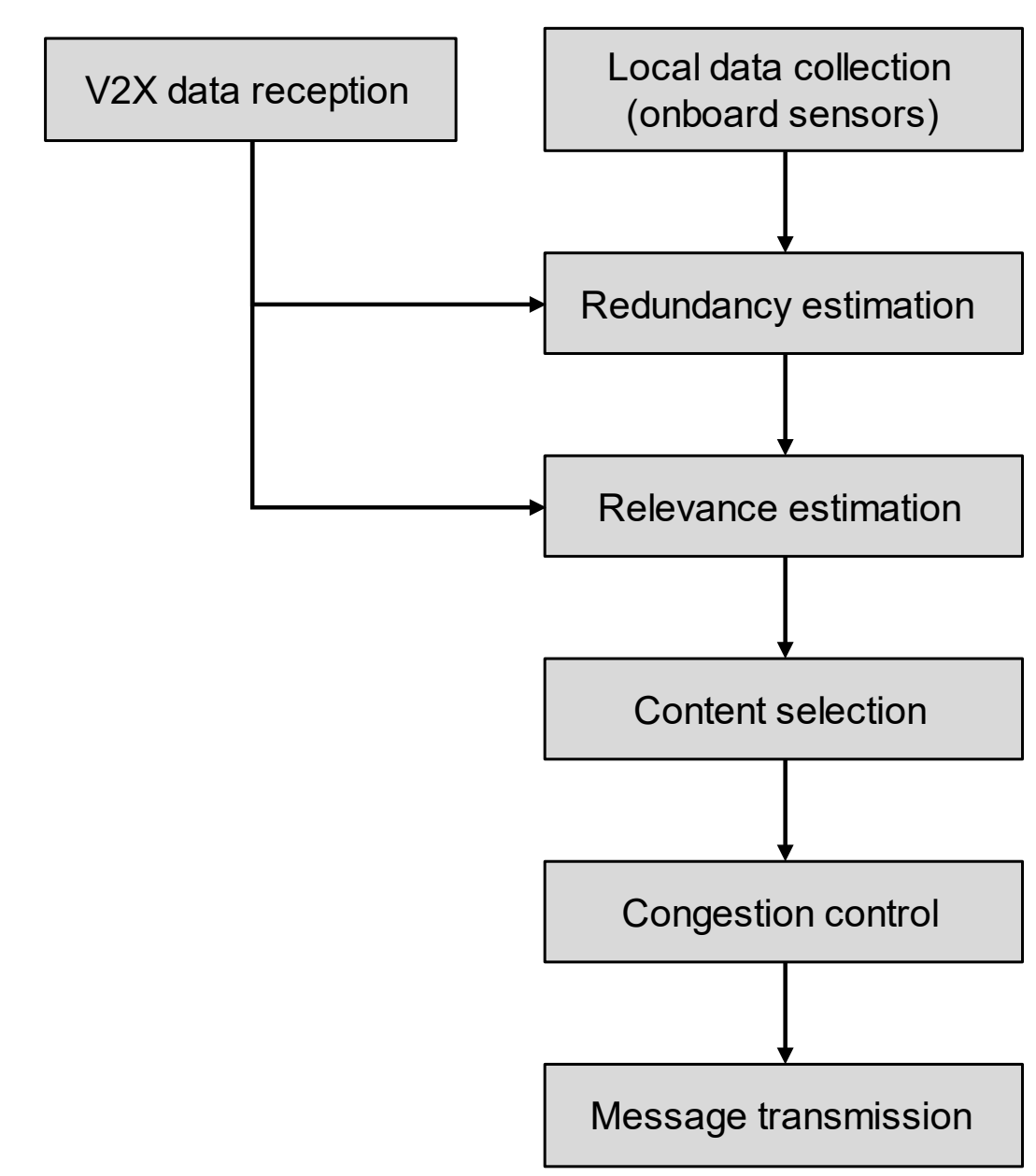

**Fig. 1.** Task-oriented V2X networks: transmitter-side block diagram.

### *A. Content-Selection Errors*

The scalability of task-oriented V2X networks strongly depends on the capability of the transmitting vehicles to correctly identify – and select – the information that is relevant to the receiving vehicles' driving. However, relevance is a context-dependent measure and context estimation can be particularly challenging due to the broadcast nature of V2X communications and the dynamism and complexity of the driving environment. A partially accurate reconstruction of the intended receivers' context can negatively affect the relevance estimation accuracy at the transmitter. As a result, a relevant piece of information can incorrectly be estimated as irrelevant and, therefore, omitted from the transmitted message – an event that can potentially cause a content-selection error. In task-oriented V2X networks, a content-selection error occurs when a message is correctly delivered at the receiver but it does not contain all relevant information available at the transmitter. A content-selection error can occur because either the redundancy or relevance of a piece of information is incorrectly estimated.

[1]In V2X, a vehicle can include only locally collected information in its messages to avoid the potential propagation of errors across the network.

[2]Note that redundant data is filtered out from the generated messages both in standardized cooperative perception [8] and in other task-oriented networks, also outside the V2X domain [10].

### B. Redundancy Estimation Errors

A redundancy estimation error occurs when the transmitter incorrectly estimates a piece of information to be already available (i.e., redundant) at an intended receiver. Due to the broadcast nature of V2X communications, a transmitting vehicle can only perform redundancy estimation by coarsely estimating the communication context of its intended receivers, i.e., by indirectly monitoring the content of the exchanged messages. As a consequence, redundancy estimation consists of probabilistically estimating whether a message overheard on the V2X network is correctly decoded by an intended receiver or not. Such a probabilistic estimation is inherently prone to errors. A redundancy estimation error is particularly critical when it involves a relevant piece of information. In this case, the transmitting vehicle can incorrectly estimate a relevant piece of information as redundant (hence, irrelevant), leading the transmitter to omit relevant information from the transmitted message and potentially causing a content-selection error. We refer to content-selection errors caused by a redundancy estimation error as *Redundancy-CSE*.

Fig. 2 depicts a cooperative perception example to analyse and illustrate the conditions that can lead to a *Redundancy-CSE*. Fig. 2 represents a T-shaped intersection with two objects ($x_1$ and $x_2$) and three communicating vehicles: $V_T$ (the transmitter), $V_O$ (the observer), and $V_R$ (the receiver). Let's assume that object $x_1$ is relevant for $V_R$ and that both $V_T$ and $V_O$ detect it. Let's further assume that $V_T$ transmits a new message, and that the content of the transmitted message includes object $x_1$. If $V_O$ correctly receives the message transmitted by $V_T$, it infers that $V_T$ has already broadcasted $x_1$ on the V2X network. If $V_O$ estimates that $V_R$ has also correctly received the message from $V_T$, it will conclude that $x_1$ is redundant (hence, irrelevant) for $V_R$ and will therefore not include it in its next transmitted message. A redundancy estimation error occurs if $V_R$ did not correctly receive the message from $V_T$ and, therefore, $x_1$ is not actually redundant for $V_R$. In this case, $x_1$ would be incorrectly estimated as irrelevant for $V_R$ and incorrectly omitted by $V_O$ from its transmitted message. If the message transmitted by $V_O$ is correctly received by $V_R$, a *Redundancy-CSE* occurs. A content-selection error directly affects the receiving vehicle, as it prevents $V_R$ from receiving any information about $x_1$.

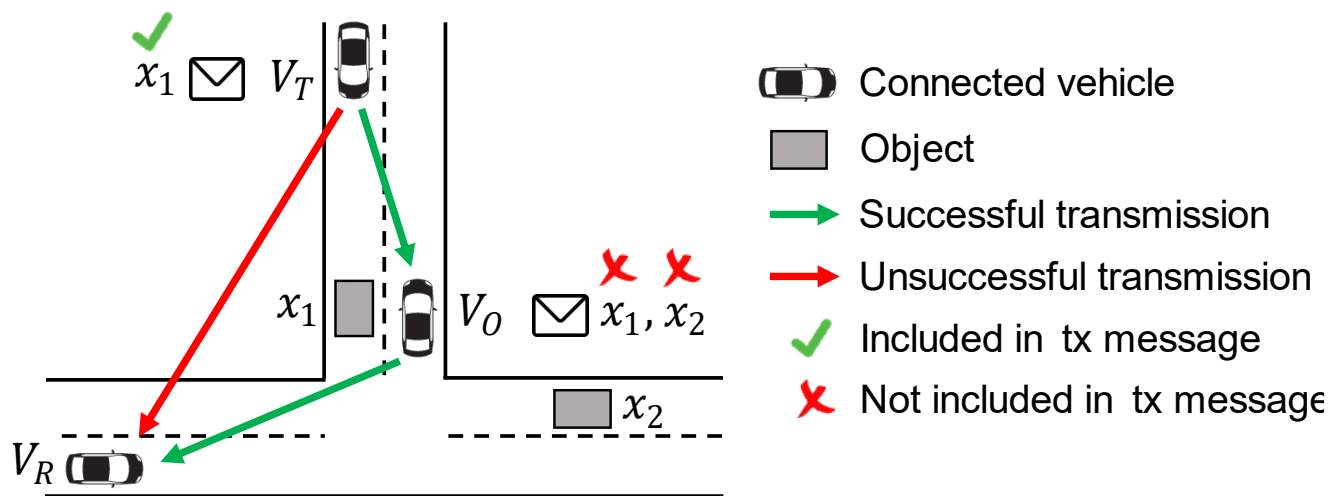

**Fig. 2.** Cooperative perception under redundancy and relevance estimation errors.

### C. Relevance Estimation Errors

A relevance estimation error occurs when the transmitter incorrectly estimates a relevant piece of non-redundant information to be irrelevant. A relevance estimation error leads the transmitter to wrongly omit relevant information from its transmitted message, causing a content-selection error. We refer to content-selection errors caused by a relevance estimation error as *Relevance-CSE*.

The relevance of a piece of non-redundant information can greatly vary depending on the context of the receiving vehicle, especially its driving context. This is particularly evident in the cooperative perception case, where the relevance of an object is strongly influenced by the current state (e.g., position, speed) and the future actions of both the receiving vehicle and the object. In the example of Fig. 2, if both $x_2$ and $V_R$ plan to go straight, $x_2$ is irrelevant for $V_R$. Conversely, if $x_2$ plans to go straight and $V_R$ plans to turn left, their future trajectories might intersect (depending also on their relative speed and distance) and, hence, $x_2$ might be relevant for $V_R$. Correctly estimating the driving context is a particularly challenging task that typically relies on probabilistic models (e.g., trajectory prediction models). These models process basic contextual information (e.g., traffic signs, road users position and speed) to achieve a better understanding of the current state and future evolution of the driving context. Due to its probabilistic nature, driving context estimation is inherently imperfect, and an inaccurate estimation of the driving context can lead to a relevance estimation error and, ultimately, to a *Relevance-CSE*.

In Fig. 2, a *Relevance-CSE* can occur if, for example, we assume that object $x_2$ will go straight at the intersection, $V_R$ will turn left and that, therefore, $x_2$ is relevant for $V_R$ as their future trajectories intersect. If $V_O$ incorrectly estimates that $V_R$ will proceed straight (inaccurate driving context estimation), $V_O$ will wrongly estimate that its locally detected object $x_2$ is irrelevant for $V_R$. $V_O$ will then omit $x_2$ from its transmitted message due to the relevance estimation error and, if the message transmitted by $V_O$ is correctly received by $V_R$, a *Relevance-CSE* occurs.

## IV. A Cooperative Perception Case Study

We analyse the impact of content-selection errors on task-oriented V2X networks considering a cooperative perception use case[3]. In cooperative perception, vehicles exchange detected objects to cooperatively overcome the limitations of their own onboard sensors and achieve a longer range and more accurate perception of the environment. In cooperative perception standards, vehicles either transmit all locally detected objects (SAE [18]) or a selected subset defined according to some predefined rules based on the objects' mobility (ETSI [8]). In task-oriented V2X networks, vehicles select the subset of locally detected objects to be transmitted based on their estimated relevance to the intended receivers.

[3]Cooperative perception is the terminology employed in ETSI (European Telecommunications Standards Institute). In SAE (Society of Automotive Engineers) terminology, cooperative perception is known as sensor data sharing.

We consider a scenario populated by $N$ connected and autonomous vehicles and $K$ exogenous variables. Exogeneous variables represent typical driving environment objects (e.g., other vehicles, pedestrians, bicycles, etc.), and are randomly distributed in the scenario following a Poisson Point Process (PPP). We denote the set of all exogenous variables in the scenario with $\mathcal{K}$ and the $k$-th exogenous variable with $x_k$, $k = 1, \dots, K$. Each vehicle $V_n$ $(n = 1, \dots, N)$ in the scenario collects exogenous variables either locally, through its onboard sensors (e.g., cameras, LiDARs, radars), or from other vehicles via V2X communications. We model the probability that a vehicle $V_n$ can locally detect an exogenous variable $x_k$ using its local sensors with the following logistic function:

$$P\left(D_{k,n}\right) = \frac{1}{1 + 0.08 * e^{0.08(D_{k,n} - 60)}}, \qquad (1)$$

where $D_{k,n}$ represents the distance between $x_k$ and $V_n$. Accordingly, each vehicle's perception range is 150 m, with the perception range representing the distance at which the probability of locally detecting an exogenous variable is equal to zero. This modelling choice is in line with the object detection probability employed in other cooperative perception studies such as [4].

Vehicles periodically generate a new message every $T$ ms and include in the generated message a subset of the locally detected exogenous variables, selected based on their estimated relevance to the intended receivers. An intended receiver is a vehicle that can potentially decode a message from the transmitter and should, therefore, be considered when selecting the content of the generated message. We assume that the transmitter estimates another vehicle as an intended receiver if it can correctly decode at least one of its last two transmitted messages. To estimate the relevance of locally collected exogenous variables, the transmitter follows the two-step process illustrated in Fig. 1. It first estimates whether the locally collected exogenous variables are redundant for the intended receivers (*Redundancy estimation* in Fig. 1). If an exogenous variable $x_k$ is estimated to be redundant for an intended receiver $V_R$, it is also deemed irrelevant for $V_R$. Redundancy is estimated at the transmitter by monitoring the exogenous variables contained in the messages previously exchanged over the V2X network. Each exogenous variable $x_k$ correctly received by the transmitter via V2X is estimated to be redundant for an intended receiver $V_R$ with probability $p_{red}$. If $V_R$ has explicitly included $x_k$ in its last transmitted message, the transmitter assumes that $x_k$ has been locally detected by $V_R$ and sets $p_{red} = 1$. Otherwise, $p_{red}$ is evaluated using one of the two following approaches:

- *Hard redundancy estimation:* the transmitter assumes that any exogenous variable it receives via V2X is also available at all intended receivers. Consequently, $p_{red}$ is fixed and equal to 1 for all intended receivers. This approach is also employed in cooperative perception studies and standards [8].
- *Soft redundancy estimation:* $p_{red}$ represents the probability that the message containing $x_k$ was correctly received by $V_R$. It is equal to the message reception probability of the V2V link between $V_R$ and the vehicle $V_T$ that reported $x_k$. $p_{red}$ depends on the network load measured at $V_R$ and the distance between $V_T$ and $V_R$.

After redundancy estimation, the transmitter estimates the relevance of non-redundant exogenous variables (*Relevance estimation* in Fig. 1). The relevance of non-redundant variables for an intended receiver $V_R$ is modelled through a contextual relevance function $f_R(\mathcal{K})$ that assigns a binary weight $w_{R,k} \in \{0,1\}$ to each non-redundant exogenous variable $x_k$. The weight $w_{R,k}$ is an indicator of the context-dependent relevance of each $x_k$ for $V_R$. Relevant variables can influence the receiver's driving and have $w_{R,k}$= 1. Irrelevant variables do not have any impact on the receiver's driving and are assigned a weight equal to zero ($w_{R,k}$ = 0). We assume that the transmitter estimates the weight $w_{R,k}$ of each $x_k$ (i.e., its relevance to $V_R$) with a relevance estimation error $\beta$. $\beta$ represents the probability that the transmitter incorrectly estimates a relevant exogenous variable as irrelevant or, vice versa, estimates an irrelevant variable as relevant. Varying degrees of $\beta$ reflect the transmitter's accuracy in estimating the intended receiver's context and, accordingly, the relevance of each detected exogenous variable $x_k$ for $V_R$. The transmitting vehicle estimates the relevance of each locally collected exogenous variable for all its intended receivers and leverages the estimated relevance to select the content of the generated messages (*Content selection* in Fig. 1). During the content-selection step, the transmitter filters out the exogenous variables that are estimated to be irrelevant for all intended receivers. It includes in the generated message only the exogenous variables that are estimated to be relevant to at least one intended receiver. Prior to its transmission, each generated message is checked by the congestion control mechanism (*Congestion control* in Fig. 1).

## V. Evaluation Scenario

We consider a Manhattan-like driving scenario covering an area of 2 x 2 km$^2$ with an exogenous variables' density equal to 750 variables/km$^2$. In this setting, the average number of exogenous variables locally detected by each vehicle is equal to 20. In a cooperative perception context, this corresponds to the average number of objects that is locally detected by a vehicle in a low-density 5 lanes highway [19] or urban scenario [20]. We assume that the maximum distance between a vehicle $V_n$ and a relevant exogenous variable is limited by the relevance range $D_{max}$. All variables beyond $D_{max}$ are assumed to be irrelevant for $V_n$. We set the relevance range $D_{max}$ equal to 400 m to capture realistic conditions in which a vehicle's driving can only be influenced by objects within a few hundred meters. We further assume that, for each vehicle $V_n$, the total number of relevant exogenous variables within the $D_{max}$ range is equal to $M$. We set $M$ = 40 to reflect the complexity of urban scenarios, where several objects – including vulnerable road users and surrounding vehicles – may influence a vehicle's driving.

Vehicles generate a new message every $T$ = 100 ms, a configuration that is valid in both ETSI [8] and SAE [18] cooperative perception standards. We set the size of an exogenous variable to 52 bytes, a setting that corresponds to the typical size of a detected object in cooperative perception [21]. We consider a 10 MHz channel bandwidth and we assume that vehicles communicate employing the QPSK-0.7 Modulation and Coding Scheme (MCS). We further assume that vehicles exchange the generated messages on the task-oriented V2X network through the C-V2X Sidelink communication technology. The performance of C-V2X sidelink communications is modelled using the analytical models released in [22]. For each transmitted message, the analytical models provide the successful message reception probability taking into account co-channel interference (i.e., collisions), signal degradation from path loss, shadowing, and fast fading, as well as the half-duplex constraints of V2X transceivers. The models express the probability of successful message reception as a function of both the transmitter-receiver distance and the Channel Busy Ratio (CBR) observed at the receiver. The CBR is an indicator of network load, as it represents the proportion of network resources that are sensed as occupied by other vehicles during the last 100 ms interval. In C-V2X sidelink, a message is sensed by a vehicle when its received signal strength surpasses a predefined detection threshold, independently of its successful decoding. The sensing process is characterized by the sensing probability model, also openly released in [22]. We implement the access layer congestion control mechanism defined in [23]. According to it, a generated message should be dropped (i.e., not transmitted) if the Channel occupancy Ratio (CR) of the transmitter is higher than a CR limit that depends on the message priority and the network load (i.e., the CBR); all CR limits and configuration parameters can be openly accessed in [23]. The CR is measured as the fraction of network resources occupied by the transmitter over a 1000 ms window.

## VI. Resilience Analysis

### *A. Redundancy Estimation Errors*

We begin our analysis by assessing the impact of *Redundancy-CSEs* on task-oriented V2X networks. We assume that the relevance estimation error $\beta$ is equal to zero and, therefore, *Relevance-CSEs* cannot occur,

Fig. 3(a) reports the content-selection error probability *P(CSE),* obtained using both hard and soft redundancy estimation, as a function of the number of communicating vehicles $N$. *P(CSE)* quantifies the probability that at least one relevant exogenous variable is not delivered to an intended receiver due to a redundancy estimation error at the transmitter. The range of considered $N$ values corresponds to network load, levels in the CBR ∈ [0, 0.7] range that are in line with other existing works [1].

Fig. 3(a) shows that, with hard redundancy estimation, *P(CSE)* steadily increases with $N$ and it can be as large as 0.15. A value of *P(CSE)* = 0.15 means that, on average, 1 every 6.6 transmitted messages does not contain all relevant information required by an intended receiver due to a redundancy estimation error at the transmitter. Such high *P(CSE)* values indicate that hard redundancy estimation may substantially reduce the amount of relevant information available at the receiving vehicles, ultimately impairing the effectiveness of task-oriented V2X networks. In contrast, Fig. 3(a) shows that soft redundancy estimation can greatly reduce the content-selection error probability and confine it below 0.05.

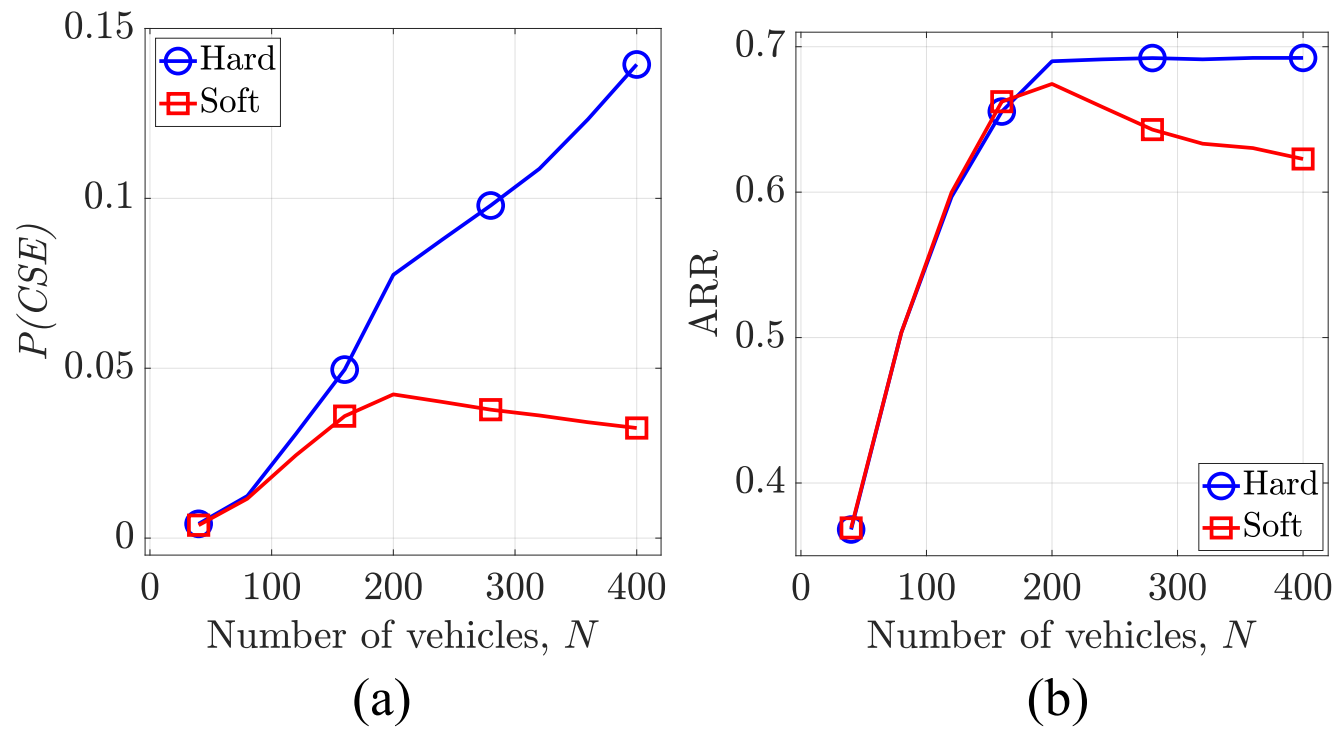

**Fig. 3.** CSE probability (a) and ARR (b) under redundancy estimation errors (*Redundancy-CSE*), reported as a function of the number of communicating vehicles, $N$.

Fig. 3(b) examines the impact of content-selection errors on the amount of relevant information delivered to each intended receiver. It depicts the Available Relevant variables Ratio (ARR) as a function of $N$. The ARR is defined as the ratio between the number of relevant exogenous variables available at each intended receiver and $M$, the total number of relevant variables within the intended receiver's relevance range. An exogenous variable is considered available if it is either locally collected by the intended receiver through its onboard sensors or received via V2X by other vehicles. Fig. 3(b) reveals that hard redundancy estimation attains larger ARR values than its soft counterpart, particularly as the number of communicating vehicles $N$ increases. Higher ARR values reflect an enhanced capability of task-oriented V2X networks to deliver larger amounts of relevant information to the intended receivers. Surprisingly, the higher content-selection error probability of hard redundancy estimation translates into a better ARR performance. The reason behind the inversion of trends observed from Fig. 3(a) to Fig. 3(b) is twofold. First, task-oriented V2X networks feature an inherent network-level resilience to content-selection errors: when a vehicle omits relevant information from a transmitted message, neighboring vehicles spontaneously intervene to transmit the omitted relevant information. Second, such network-level resilience is stronger when hard redundancy estimation is employed. Soft redundancy estimation wastes a large amount of communication resources and limits the vehicles' capability to recover each other's content-selection errors.

To illustrate the network-level resilience of task-oriented V2X networks against content-selection errors, Fig. 4(a) reports the CSE recovery attempt probability, denoted as $P_{att}(CSE)$. A CSE recovery attempt occurs when a vehicle includes in its transmitted message a piece of relevant information that was previously omitted by one of its neighbors, thereby causing a CSE. Fig. 4(a) shows that both redundancy estimation techniques achieve $P_{att}(CSE)$ values larger than 0.5 across all considered vehicular densities ($N$). This means that, on average, more than 50% of content-selection errors are subject to recovery attempts. Such large $P_{att}(CSE)$ values explain the resilience against content-selection errors observed in Fig. 3. Fig. 4(b) completes the CSE recovery analysis by focusing on the CSE successful recovery probability, denoted as $P_{succ}(CSE)$. A recovery attempt is successful when it can be correctly delivered to the intended receiver that did not previously receive the relevant information due to the CSE. The $P_{succ}(CSE)$ curves exhibit similar trends, but with smaller absolute values, compared to the $P_{att}(CSE)$ curves reported in Fig. 4(a). This is the case since, by definition, a successful recovery is conditioned on the correct reception of the recovery attempt. Nevertheless, $P_{succ}(CSE)$ remains above 0.5, regardless of the considered vehicular density ($N$) or redundancy estimation technique. This indicates that, on average, the majority of CSEs can be successfully recovered in task-oriented V2X networks, and the intended receivers can eventually receive previously omitted relevant information. We numerically verified that this is case also for other configurations of the main system parameters (e.g., exogenous variables density, $D_{max}$ and $M$).

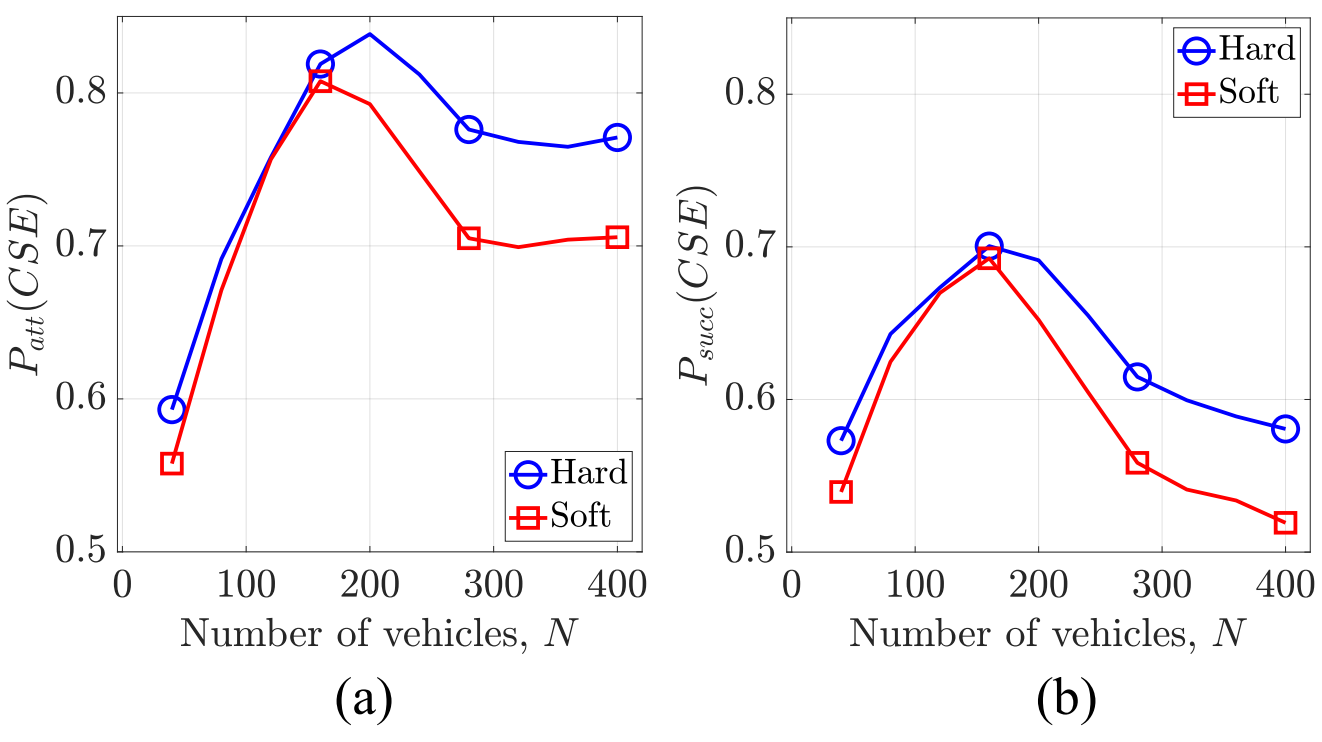

**Fig. 4.** CSE recovery attempt (a) and successful recovery (b) probability under redundancy estimation errors (*Redundancy-CSE)*, reported as a function of $N$.

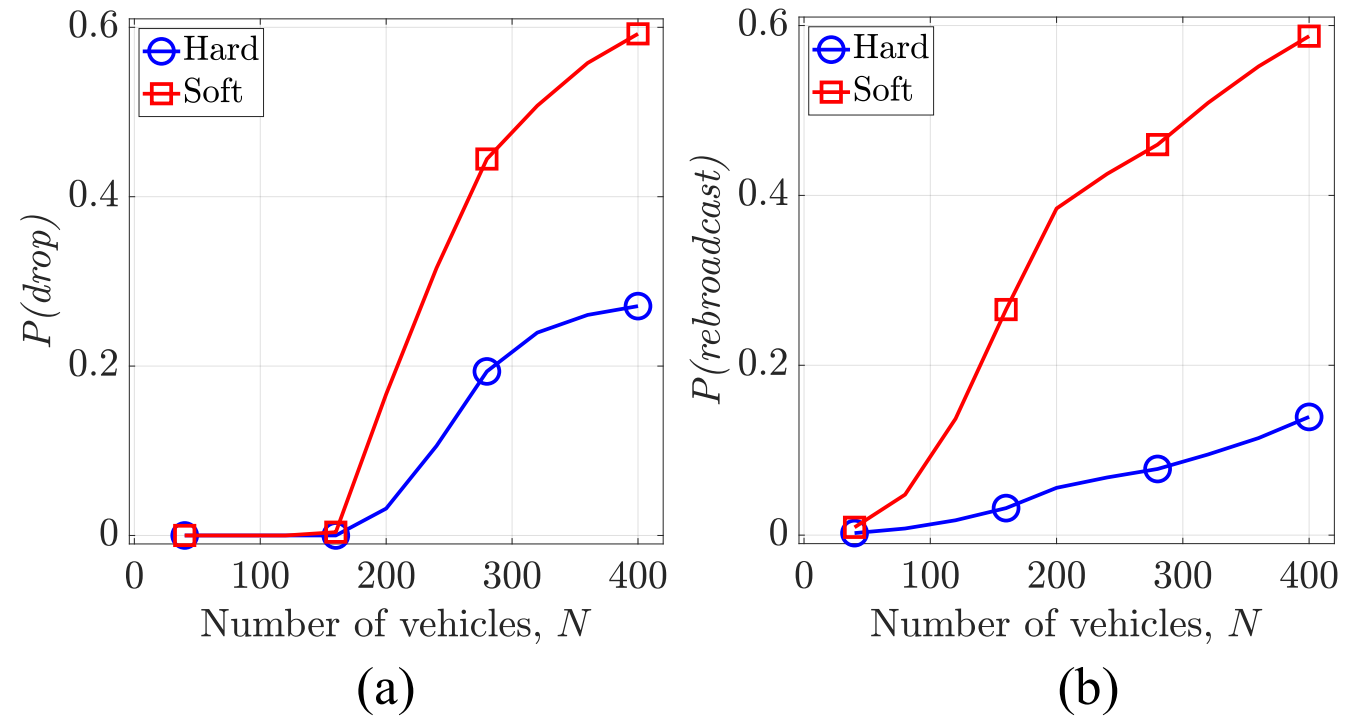

**Fig. 5.** Drop (a) and rebroadcast (b) probability under redundancy estimation errors (*Redundancy-CSE)*, reported as a function of $N$.

The $P_{att}(CSE)$ and $P_{succ}(CSE)$ analysis sheds light on the inherent resilience of task-oriented V2X networks. Vehicles spontaneously attempt to compensate each other's CSEs and collaboratively provide the intended receivers with all relevant driving environment information. This resilience naturally stems from the likelihood of having multiple vehicles that locally detect the same relevant exogenous variables and are therefore able to recover each other's content-selection errors. For this reason, $P_{att}(CSE)$ and $P_{succ}(CSE)$ increase with $N$ at small-medium vehicular densities ($N < 200$). An increase in the number of communicating vehicles augments the likelihood of having multiple vehicles locally detecting the same relevant information, thereby leading to a stronger resilience. However, the positive correlation between $N$ and both $P_{att}(CSE)$ and $P_{succ}(CSE)$ is no longer observed at medium-high vehicular densities ($N > 200$), where the curves exhibit a decreasing trend. In this range of $N$ values, Fig. 4(a) and Fig. 4(b) also show that hard redundancy estimation leads to a stronger network-level resilience, as it yields better $P_{att}(CSE)$ and $P_{succ}(CSE)$ values compared to soft redundancy estimation. On average, the amount of successfully recovered content-selection errors is 10% larger with hard redundancy estimation.

In the $N > 200$ region, the reduction in the observed $P_{att}(CSE)$ and $P_{succ}(CSE)$ values, as well as the superiority of hard redundancy estimation, stem from the impact of the congestion control mechanism. This impact is measured by the drop probability, denoted as *P(drop)*, which represents the likelihood that a generated message is dropped (i.e., discarded) by the congestion control mechanism to maintain the network load (i.e., the CBR) below a predefined level (see Fig. 1). Fig. 5(a) reports *P(drop)* as a function of $N$. As long as $N < 200$, *P(drop)* = 0 and vehicles do not face any communication constraints, i.e., congestion control does not limit the amount of information that each vehicle can transmit. Therefore, a larger number of communicating vehicles (i.e., larger $N$) directly translates into a stronger network-level resilience, as reflected by the increasing $P_{att}(CSE)$ and $P_{succ}(CSE)$ values. Conversely, when $N < 200$, the network load increases beyond the predefined limits and congestion control intervenes (by dropping packets) to limit the amount of information transmitted by each vehicle. As a result, *P(drop)* > 0 and the vehicles' capability to attempt a CSE recovery is reduced, thereby leading to smaller $P_{att}(CSE)$ and $P_{succ}(CSE)$ values. Fig. 5(a) also highlights that soft redundancy estimation significantly increases the drop probability compared to hard redundancy estimation. It leads to up to a 2x increase in *P(drop)* at high vehicular densities. A larger *P(drop)* not only limits the CSE recovery probability (see Fig. 4(b)), but also constraints the capability of each vehicle to transmit relevant information to its intended receivers.

Soft redundancy estimation adopts a probabilistic approach that lowers the likelihood of estimating information as redundant. While this reduces the redundancy estimation error probability and, in turn, the content-selection error probability (see Fig. 3(a)), it also increases the likelihood that an exogenous variable is redundantly broadcasted by multiple vehicles, thereby augmenting the network load and the drop probability. This tradeoff is illustrated in Fig. 5(b), which reports the rebroadcast probability, *P(rebroadcast)*, as a function of $N$. *P(rebroadcast)* is defined as the probability that a transmitter rebroadcasts an exogenous variable already reported by another vehicle. Fig. 5(b) shows that soft redundancy estimation significantly increases the rebroadcast probability compared to hard redundancy estimation. As the vehicular density grows, *P(rebroadcast)* remains below 0.2 with hard redundancy estimation, while it can reach up to 0.6 if soft redundancy estimation is employed – a more than threefold increase. However, the largest fraction of the additional rebroadcasts performed by soft redundancy mitigation is useless for the intended receivers: on average, the 80% of rebroadcasted variables is already available (i.e., redundant) at the intended receiver. Redundant information does not improve the intended receiver's perception and understanding of the driving environment and is considered irrelevant. These additional rebroadcasts waste a considerable amount of communication resources, significantly increasing *P(drop)* and eventually limiting the vehicles' capability to recover each other's CSEs. Ultimately, they reduce the network-level resilience against content-selection errors (see Fig. 4(b)) compared to hard redundancy estimation, and are responsible for the ARR reduction observed in Fig. 3(b) for $N > 200$.

*B. Redundancy and Relevance Estimation Errors*

Based on the previous section results, we now analyse the resilience of task-oriented V2X networks against both *Redundancy-CSEs* and *Relevance-CSEs* focusing on hard redundancy estimation. Fig. 6(a) reports the content-selection error probability, *P(CSE)*, obtained under different relevance estimation error probabilities, $\beta$. Fig. 6(a) shows that $\beta$ has a strong impact on the content-selection error probability. When $\beta = 0.8$, the content-selection error probability *P(CSE)* is equal to 0.4, indicating that a transmitting vehicle omits relevant information once every 2.5 generated messages. This is a substantial increase compared to the $\beta = 0$ case where relevance estimation is ideal and *P(CSE)* is confined below 0.15.

The impact of relevance estimation errors on task-oriented V2X networks is analysed in Fig. 6(b), which reports the ARR obtained for different $\beta$ values. The comparison between Fig. 6(a) and Fig. 6(b) shows that a larger content-selection error probability translates into smaller ARR values, i.e., smaller amounts of relevant information delivered (hence, available) at the receiving vehicles. In addition, the comparison reveals that the impact of $\beta$ is much smaller in the ARR case as long as $\beta \leq 0.4$, especially at large $N$ values. For example, the ARR performance gap between the $\beta = 0$ and $\beta = 0.4$ scenarios is small when $N > 300$, despite a substantial difference in terms of content-selection error probability (see Fig. 6(a)). At large vehicular densities, the likelihood of having multiple vehicles detecting the same relevant information increases. As a result, the resilience of task-oriented V2X networks guarantees the consistent delivery of relevant information (i.e., a stable ARR performance) also in high *P(CSE)* scenarios. Note that when $\beta = 0.4$, 1 every 3.3 transmitted messages does not contain all relevant information required by the intended receivers. Yet, the task-oriented V2X network is able to guarantee ARR values very close to the $\beta = 0$ scenario.

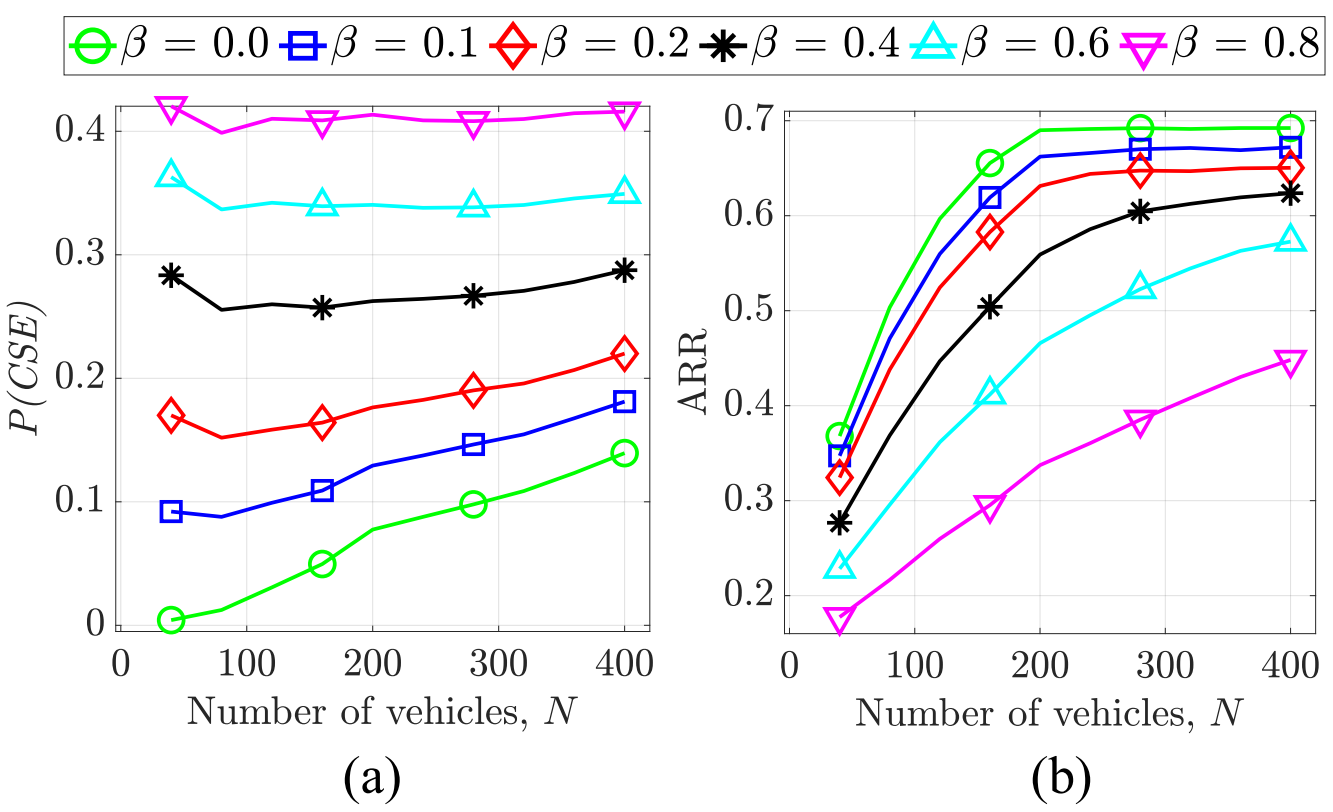

**Fig. 6.** CSE probability (a) and ARR (b) under redundancy (*Redundancy-CSE)* and relevance estimation errors (*Relevance-CSE)*, reported as a function of $N$.

Yet, the task-oriented V2X network resilience can be compromised if the relevance estimation error becomes excessively large. This is particularly visible when $\beta = 0.8$. In this scenario, vehicles are no longer able to recover for each other's content-selection errors and, as a consequence, the ARR performance substantially drops. A large relevance estimation error $\beta$ has two distinct effects: (i) it increases the probability that the transmitter omits relevant information from its message (see Fig. 6(a)); and (ii) it prevents neighbouring vehicles from correctly identifying – and transmitting – the omitted relevant information and compensate for the content-selection error. The latter effect is quite evident in Fig. 7(a), which reports the CSE successful recovery probability, $P_{succ}$*(CSE)*, obtained for different $\beta$ scenarios. This figure shows that vehicles can collaboratively compensate for each other's content-selection errors and achieve the same $P_{succ}$*(CSE)* performance at large vehicular densities ($N > 300$) as long as $\beta \leq 0.4$[4]. This is not the case for the $\beta = 0.6$ and $\beta = 0.8$ scenarios, where smaller $P_{succ}$*(CSE)* values are observed across all vehicular densities.

Fig. 7(a) also highlights that a larger vehicular density (i.e., a larger $N$) always has a positive impact on the resilience of the task-oriented V2X network, regardless of the $\beta$ setting. In all curves, $P_{succ}$*(CSE)* increases with $N$ until the congestion control mechanism intervenes to limit the network load (i.e., for $N > 200$). Moreover, Fig. 7(a) reveals that all $P_{succ}$*(CSE)* curves exhibit similar trends at small-medium vehicular densities ($N < 200$), except for the $\beta = 0$ case. When $\beta > 0$, $P_{succ}$*(CSE)* grows starting from negligible values. Conversely, when $\beta = 0$, $P_{succ}$*(CSE)* already exceeds 0.5 also at small vehicular densities. These different trends stem from two main factors. First, *Redundancy-CSEs* and *Relevance-CSEs* occur under different conditions. A *Redundancy-CSE* can occur only if at least two neighbouring vehicles detect the same relevant variable and one of them includes it in its last message (see Fig. 2). Hence, when a *Redundancy-CSE* occurs, the likelihood that another neighbouring vehicle has also detected the same relevant information – and can recover the CSE – is high, even at small vehicular densities. Conversely, a *Relevance-CSE* can occur even if a relevant variable is detected by a single vehicle (the transmitter). The occurrence of a *Relevance-CSE* exclusively depends on the transmitter's capabilities to accurately estimate the context of its intended receivers. The likelihood that another vehicle has detected the same relevant information – and can recover the *Relevance-CSE* – strongly depends on the vehicular density ($N$). Second, when $\beta = 0$, all content-selection errors are due to a redundancy estimation error and, therefore, the network-level resilience exclusively depends on the vehicles' capability to recover *Redundancy-CSEs.* Instead, at small-medium vehicular densities, the majority of content-selection errors is due to a relevance estimation error (*Relevance-CSEs*) when $\beta > 0$, as shown in Fig. 7(b). Fig. 7(b) depicts the fraction of *Redundancy-CSEs*, calculated with respect to the total number of CSEs, as a function of $N$. Fig. 7(b) shows that the fraction of *Redundancy-CSEs* depends on the number of communicating vehicles, $N$, and that it is not negligible across all $\beta$ configurations at medium-large vehicular densities ($N > 200$), where the network-level resilience is stronger (see Fig. 7(a)). This indicates that the resilience of task-oriented V2X networks stems from the capability of the transmitting vehicles to collaboratively recover both *Redundancy-CSEs* and *Relevance-CSEs.* We numerically verified the results and trends reported in this section under an exhaustive set of system configurations, i.e., for varying exogenous variables density, $D_{max}$ and $M$ configurations.

[4] Scenarios with $\beta \leq 0.4$ do not have the same ARR performance in Fig. 6(b) despite having the same $P_{succ}$*(CSE)* performance in Fig. 7(a) because the total number of CSEs is larger in higher $\beta$ settings as shown by the *P(CSE)* values reported in Fig. 6(a).

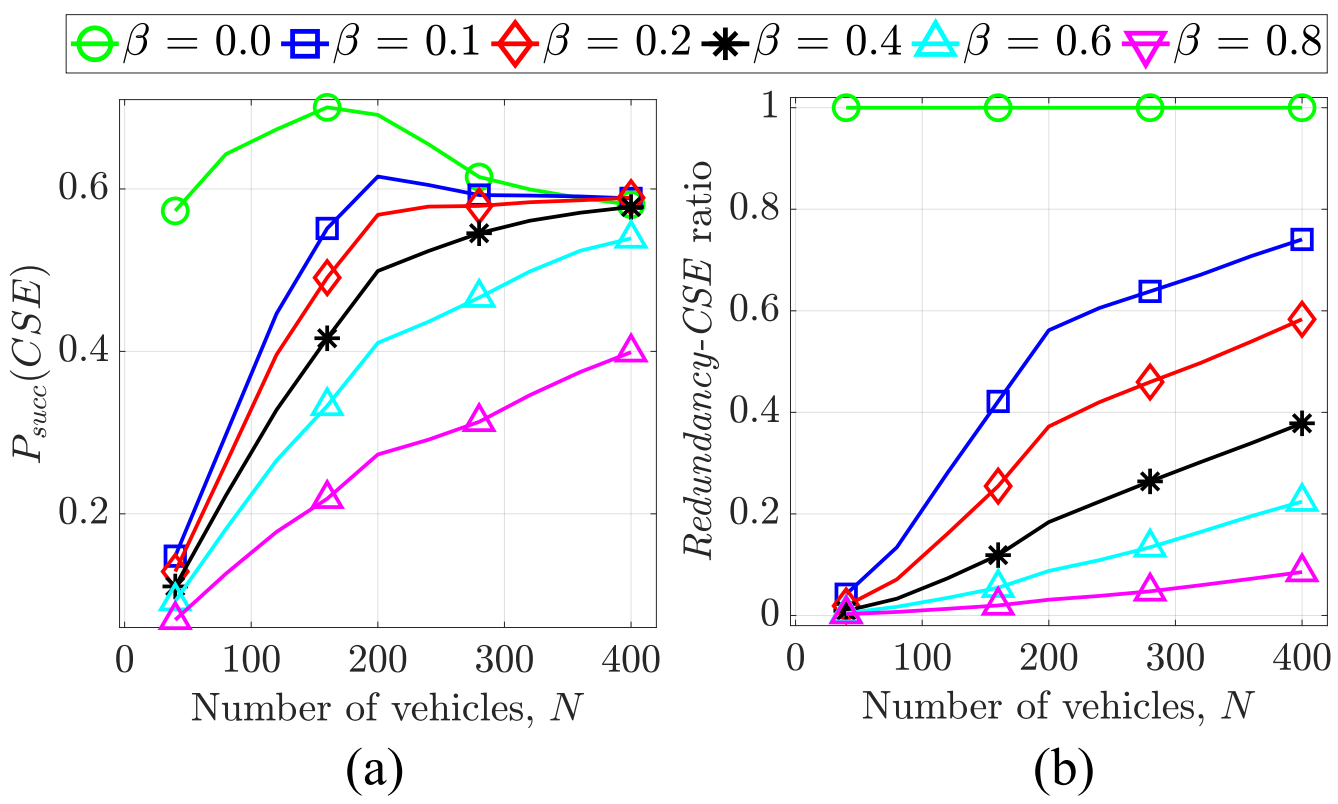

**Fig. 7.** CSE successful recovery probability (a) and *Redundancy-CSE* ratio (b) under redundancy (*Redundancy-CSE)* and relevance estimation errors (*Relevance-CSE)*, reported as a function of $N$.

## VII. Conclusions

Task-oriented V2X networks have the potential to greatly improve communication efficiency and address future scalability challenges. Their effectiveness depends on the ability of the transmitting vehicles to identify and transmit only the relevant information required by the receivers – a critical yet largely unexplored challenge. In task-oriented V2X networks, an inaccurate redundancy or relevance estimation can lead to content-selection errors that can ultimately lead to the transmission of incomplete information, potentially reducing the amount of relevant information delivered to the receivers and impairing their situational awareness.

This work analyses the impact of content-selection errors on task-oriented V2X networks. Our analysis quantifies the occurrence probability of content-selection errors considering different redundancy estimation techniques (hard and soft) and relevance estimation error probabilities. Moreover, it sheds light on the conditions under which a redundancy or relevance estimation error can occur and, ultimately, lead to a content-selection error. Notably, our study reveals that task-oriented V2X networks feature an inherent resilience to content-selection errors that guarantees the consistent delivery of relevant information to the receiving vehicles also in scenarios with a non-negligible content-selection error probability. This inherent resilience relaxes the need for extremely accurate redundancy and relevance estimation techniques, facilitating the design and deployment of future task-oriented V2X networks. The obtained results show that the resilience augments for increasing vehicular densities, as long as congestion control mechanisms do not intervene to limit the network load. We demonstrate that this resilience stems from the network-level capability to compensate for content-selection errors due to both redundancy and relevance estimation errors. In the V2X domain, this capability emerges from the likelihood of having multiple vehicles detecting the same (relevant) information with their sensors and, therefore, being able to recover each other's content-selection errors. In general, this condition can be potentially encountered in other types of task-oriented networks where, for example, multiple sensing devices monitor an overlapping portion of the environment and curate the content of their messages based on the task-related requirements of a common set of intended receivers (e.g., edge servers). The broad scope of our findings extends the value of our analysis, providing valuable insights for the design of task-oriented networks within the V2X domain and beyond.